\documentclass{article}
\usepackage{graphicx}
\usepackage{amssymb}

\begin{document}

\title{Scheduling Links for Heavy Traffic on Interfering Routes in Wireless Mesh
Networks}

\author{Fabio~R.~J.~Vieira$^{1,2,}$\thanks{Corresponding author (fjimenez@cos.ufrj.br).}\\
Jos\'e~F.~de~Rezende$^3$\\
Valmir~C.~Barbosa$^1$\\
Serge~Fdida$^2$\\
\\
$^1$Programa de Engenharia de Sistemas e Computa\c c\~ao, COPPE\\
Universidade Federal do Rio de Janeiro\\
Caixa Postal 68511, 21941-972 Rio de Janeiro - RJ, Brazil\\
$^2$Laboratoire d'Informatique de Paris 6\\
4, Place Jussieu, 75252 Paris Cedex 05, France\\
$^3$Programa de Engenharia El\'etrica, COPPE\\
Universidade Federal do Rio de Janeiro\\
Caixa Postal 68504, 21941-972 Rio de Janeiro - RJ, Brazil}

\date{}

\maketitle

\begin{abstract}
We consider wireless mesh networks and the problem of scheduling the links of a
given set of routes under the assumption of a heavy-traffic pattern. We assume
some TDMA protocol provides a background of synchronized time slots and seek to
schedule the routes' links to maximize the number of packets that get delivered
to their destinations per time slot. Our approach is to construct an undirected
graph $G$ and to heuristically obtain node multicolorings for $G$ that can be
turned into efficient link schedules. In $G$ each node represents a link to be
scheduled and the edges are set up to represent every possible interference for
any given set of interference assumptions. We present two multicoloring-based
heuristics and study their performance through extensive simulations. One of the
two heuristics is based on relaxing the notion of a node multicoloring by
dynamically exploiting the availability of communication opportunities that
would otherwise be wasted. We have found that, as a consequence, its performance
is significantly superior to the other's.

\end{abstract}

\newpage
\section{Introduction}\label{sec:intro}

Owing to their numerous advantages, wireless mesh networks (WMNs) constitute a
promising solution for community networks and for providing last-mile
connections to Internet users \cite{aww05,bvb05,nnshwa07}. However, like all
wireless networks WMNs suffer from the problem of decreased capacity as they
become denser, since in this case attempting simultaneous transmissions causes
interference to increase significantly \cite{gk00,sgpsbb07}. One common solution
to reduce interference is to adopt some contention-free TDMA protocol
\cite{dea06} and schedule simultaneous transmissions for activation only if they
do not interfere with one another. Doing this while maximizing some measure of
network usage and guaranteeing that all links are given a fair treatment
normally translates into a complicated optimization problem, one that
unfortunately is NP-hard \cite{bbkmt04}.

This scheduling problem has been formulated in a great variety of manners and
has received considerable attention in the literature. Prominent studies include
some that seek to calculate the capacity of the network \cite{gk00,gwhw09},
others whose goal is the study of the time complexity associated with the
resulting schedules \cite{mwz06}, and still others that aim at scheduling
transmissions in order to achieve as much of the network's capacity as possible
\cite{cs03,abl05,wwlsf06,gdp08,hl08,wdjhl08,smrdb09,xt09,wg09}. One common
thread through most the latter is that, having adopted a graph representation of
the network and of how the various transmissions can interfere with one another,
a solution is sought through some form of graph coloring. More often than not
the transmissions to be scheduled are represented by the graph's nodes and then
node coloring, through the abstraction of an independent set to represent the
transmissions that can take place simultaneously, is used. But sometimes it is
the graph's edges that stand for transmissions, in which case edge coloring is
used, building on the abstraction of matchings to represent simultaneity
\cite{bm08}.

Here we consider a variation of the problem which, to the best of our knowledge,
is novel both in its formulation and in the solution type we propose. We start
by assuming a WMN comprising single-channel, single-radio nodes and for which a
set of origin-to-destination routes has already been determined, and consider
the following question. Should there be an infinite supply of packets at each
origin to be delivered to the corresponding destination in the FIFO order, and
should all nodes in the network be endowed with only a finite number of buffers
for the temporary storage of in-transit packets, how can transmissions be
scheduled to maximize the number of packets that get delivered to the
destinations per TDMA slot without ever stalling a transmission, by lack of
buffering space, whenever it is scheduled? This question addresses issues that
lie at the core of successfully designing WMNs and their routing protocols,
since it seeks to tackle the problem of transmission interference when the
network is maximally strained. The solution we propose is, like in so many of
the approaches mentioned above, based on coloring a graph's nodes. Unlike them,
however, we use node multicolorings instead \cite{s76}, which are more general
and for this reason allow for a more suitable formulation of the optimization
problem to be solved.

Given the origin-to-destination routes (or paths) to be used, we begin in
Section~\ref{sec:form} with a precise definition of a schedule and a precise
formulation of the problem. We also show, through an example, that had the
problem been formulated for network-capacity maximization, a conflict with the
requirement of finite buffering might arise. Then we move, in
Section~\ref{sec:transf}, to the specification of the undirected graph that
underlies our algorithm's operation. One assumption in that section, and
therefore throughout most of the paper, is that the communication and
interference radii are the same for the WMN at hand. Moreover, we also assume
that the tenets of the protocol-based interference model \cite{br03,shlk09},
including the possibility of bidirectional communication in each transmission,
are in effect. In Section~\ref{sec:multic} we guide the reader through various
multicoloring possibilities, which culminates in Section~\ref{sec:ser} with a
preliminary method for scheduling, borrowed from the field of resource sharing
\cite{b96}. Improving on this preliminary method with the goals of the problem
formulated in Section~\ref{sec:form} in mind finally yields our proposal in
Section~\ref{sec:sera}. This proposal, essentially, stems from a slight
relaxation of the notion of a node multicoloring. The subsequent two sections
are dedicated to the presentation of computational results, with the methodology
laid down in Section~\ref{sec:methods} and the results proper in
Section~\ref{sec:results}. Discussion follows in Section~\ref{sec:disc} and we
close in Section~\ref{sec:concl}.

\section{Problem formulation}\label{sec:form}

We consider a collection $\mathcal{P}_1,\mathcal{P}_2,\ldots,\mathcal{P}_P$ of
simple directed paths (i.e., directed paths that visit no node twice), each
having at least two nodes (a source and a destination). These paths' sets of
nodes are $X_1,X_2,\ldots,X_P$, respectively, not necessarily disjoint from one
another, and we let $X=\bigcup_{p=1}^PX_p$. Their sets of edges are
$Y_1,Y_2,\ldots,Y_P$ and we assume that, for $p\neq q$, a member of $Y_p$ and
one of $Y_q$ are distinguishable from each other even if they join the same two
nodes in the same direction. Letting $Y=\bigcup_{p=1}^PY_p$, we then see that
$Y$ may contain more than one edge joining the same two nodes in the same
direction (parallel edges) or in opposing directions (antiparallel edges).

\begin{figure}[t]
\centering
\scalebox{0.75}{\includegraphics{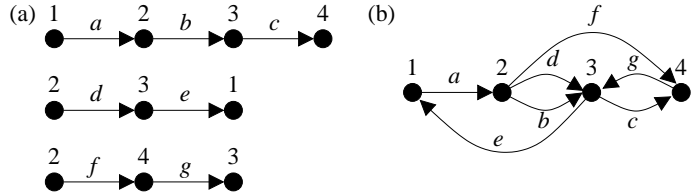}}
\caption{A set of $P=3$ directed paths (a) and the resulting directed multigraph
$D$ (b).}
\label{fig1}
\end{figure}

Our discussion begins with the definition of the directed multigraph
$D=(X,Y)$, where all $P$ directed paths are represented without sharing any
directed edges among them. An example is shown in Figure~\ref{fig1}. We take
$D$ to be representative of a wireless network operating under some TDMA
protocol. In this network, each of paths
$\mathcal{P}_1,\mathcal{P}_2,\ldots,\mathcal{P}_P$ is to transmit an unbounded
sequence of packets from its source to its destination. Such transmissions are
to occur without contention, meaning that whenever an edge is scheduled to
transmit in a given time slot no other edge that can possibly interfere with
that transmission is to be scheduled at the same time slot. We assume that each
transmission sends at most one packet across the edge in question (more
specifically, it sends exactly one packet if there is at least one to be sent
but does nothing otherwise). We also assume that each transmission may involve
the need for bidirectional communication for error control.

We call a schedule any finite sequence
$\mathcal{S}=\langle S_0,S_1,\ldots,S_{L-1}\rangle$ such that
$S_\ell\subseteq Y$ for $0\le\ell\le L-1$, provided
$\bigcup_{\ell=0}^{L-1}S_\ell=Y$ and moreover no two concurrent transmissions on
edges of the same $S_\ell$ can interfere with each other. To schedule the
transmissions according to $\mathcal{S}$ is to cycle through the edge sets
$S_0,S_1,\ldots,S_{L-1}$, indefinitely and in this order, letting all edges in
the same set transmit in the same time slot whenever that set is reached along
the cycling. Given $\mathcal{S}$, we let $\mathrm{length}(\mathcal{S})=L$ and
denote by $\mathrm{delivered}(\mathcal{S})$ the number of packets that can get
delivered to all paths' destinations during a single repetition of $\mathcal{S}$
in the long run (i.e., in the limit as the number of repetitions grows without
bound). Of course, $\mathrm{delivered}(\mathcal{S})$ is bounded from above by
the number of times the $P$ paths' terminal edges (those leading directly to a
destination node) appear in $\mathcal{S}$ altogether.

Before we use these two quantities to define the optimization problem of finding
a suitable schedule for $D$, we must recognize that our focus on the
source-to-destination packet flows on the paths
$\mathcal{P}_1,\mathcal{P}_2,\ldots,\mathcal{P}_P$ carries with it the inherent
constraint that the nodes' capacity to buffer in-transit packets cannot be
allowed to grow unbounded. We then adopt an upper bound $B$ on the number of
in-transit packets that a node can store for each of the paths (at most $P$)
that go through it. However, there is still a decision to be made regarding the
effect of such a bound on the transmission of packets. One possibility would be
to impose that, when it is an edge's turn to transmit it does so if and only if
there is a packet to transmit and, moreover, there is room to store that packet
if it is received as an in-transit packet. Another possibility, one that seeks
to never stall a transmission by lack of a buffer to store the packet at the
next intermediate node, is to only admit schedules that automatically rule out
the occurrence of such a transmission. We adopt the latter alternative.

The following, then, is how we formulate our scheduling problem on $D$. Find a
schedule $\mathcal{S}$ that maximizes the throughput
\begin{equation}
T(\mathcal{S})=
\frac{\mathrm{delivered}(\mathcal{S})}{\mathrm{length}(\mathcal{S})},
\label{eq:throughput}
\end{equation}
subject to the following two constraints:
\begin{itemize}
\item[C1.] Every node can store up to $B$ in-transit packets for each of the
source-to-destination paths that go through it.
\item[C2.] Whenever an edge is scheduled for transmission in a time slot and a
packet is available to be transmitted, if the edge is not the last one on its
source-to-destination path then there has to be room for the packet to be stored
after it is transmitted.
\end{itemize}

\subsection{Scheduling for maximum network usage}

Before proceeding, recall that, as mentioned in Section~\ref{sec:intro}, the
most commonly solved problem regarding the selection of a schedule $\mathcal{S}$
is not the one we just posed, but rather the problem of maximizing network
usage. In terms of our notation, this problem requires that we find a schedule
that maximizes
\begin{equation}
U(\mathcal{S})=
\frac{\sum_{\ell=0}^{L-1}\vert S_\ell\vert}{\mathrm{length}(\mathcal{S})}
\end{equation}
without any constraints other than those that already participate in the
definition of a schedule.

\begin{figure}[t]
\centering
\scalebox{0.75}{\includegraphics{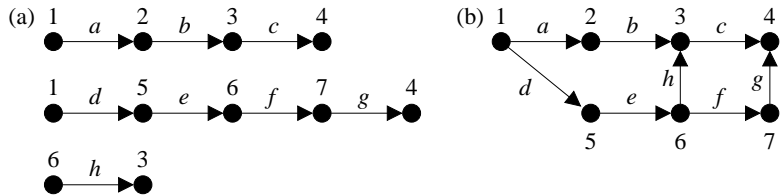}}
\caption{A set of $P=3$ directed paths (a) and the resulting directed multigraph
$D$ (b). Using the schedule $\mathcal{S}$ such that $S_0=\{a,f\}$,
$S_1=\{c,d\}$, $S_2=\{b\}$, $S_3=\{e\}$, $S_4=\{a,g\}$, and $S_5=\{h\}$ causes
unbounded packet accumulation at node $2$ when constraint~C2 is in effect, thus
violating constraint~C1. Enforcing constraint~C1 for some value of $B$ causes
constraint~C2 to be violated.}
\label{fig2}
\end{figure}

It is a simple matter to verify that solutions to this problem often fail to
respect constraints~C1 and~C2 of our formulation. This is exemplified in
Figure~\ref{fig2}.

\section{Graph transformation}\label{sec:transf}

\begin{figure}[t]
\centering
\scalebox{0.75}{\includegraphics{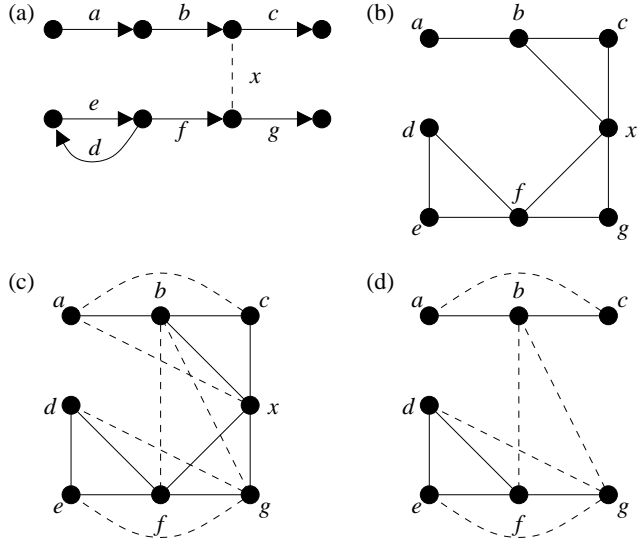}}
\caption{The graph-transformation process. We start with the directed multigraph
$D$ (a), to which the node pair labeled $x$ is added as a dashed line to
indicate the existence of interference that is not internal to any of the
initial $P$ paths. Panel (b) contains the undirected graph $G$ as it stands
after Step~2.ii. Panels (c) and (d) show $G$ past Steps~2.iii and~2.iv,
respectively. In these two panels, dashed lines are used to represent the edges
added in Step~2.iii.}
\label{fig3}
\end{figure}

We wish to address the problem of optimizing $T(\mathcal{S})$ exclusively in
terms of some underlying graph. Clearly, though, the directed multigraph $D$ is
not a good candidate for this, since it does not embody any representation of
how concurrent transmissions on its edges can interfere with one another. Our
first step is then to transform $D$ into some more suitable entity, which will
be the undirected graph $G=(N,E)$ defined as follows:
\begin{enumerate}
\item The node set $N$ of $G$ is the edge set $Y$ of $D$. In other words, $G$
has a node for every edge of $D$. Since $D$ is a multigraph, a same pair of
nodes $i,j\in X$ such that $(i,j)\in Y$ or $(j,i)\in Y$ may appear more than
once as a member of $N$.
\item The edge set $E$ of $G$ is obtained along the following four steps:
\begin{enumerate}
\item[i.] Enlarge $N$ by including in it all node pairs of $D$ that do not
correspond to edges on any of the $P$ source-to-destination paths but
nevertheless reflect that each node in the pair is within the interference
radius of the other. We refer to these extra members of $N$ as temporary nodes.
\item[ii.] Connect any two nodes in $N$ by an edge if, when regarded as node
pairs from $D$, they share at least one of the nodes of $D$. In other words, if
each of the two pairs $i,j\in X$ and $k,l\in X$ corresponds to a node of $G$ (by
virtue of either constituting an edge of $D$ or being a temporary node), then
the two get connected by an edge in $G$ if at least one of $i=k$, $i=l$, $j=k$,
or $j=l$ holds.
\item[iii.] Connect any two nodes in $N$ by an edge if, after the previous
steps, the distance between them is $2$.
\item[iv.] Eliminate all temporary nodes from $N$ and all edges from $E$ that
touch them.
\end{enumerate}
Together, these four steps amount to using $G$ to represent every possible
interference that may arise under the assumptions of the protocol-based model
when communication is bidirectional. Graph $G$ is also known as a distance-$2$
graph relative to $D$ \cite{bbkmt04}. The entire transformation process, from
the set of $P$ paths through graph $G$, is illustrated in Figure~\ref{fig3}.
\end{enumerate}

It follows from this definition of $G$ that any group of nodes corresponding to
parallel or antiparallel edges in $D$ are a clique (a completely connected
subgraph) of $G$. Similarly, every group of three consecutive edges on any of
the paths $\mathcal{P}_1,\mathcal{P}_2,\ldots,\mathcal{P}_P$ corresponds to a
three-node clique in $G$. As we discuss in Section~\ref{sec:disc}, these and
other cliques are related to how large $T(\mathcal{S})$ can be under one of the
scheduling methods we introduce.

It is also worth noting that Steps~1 and~2 above are easily adaptable to
modifications in any of the assumptions we made. These include the assumptions
that the communication and interference radii are the same and that
communication is bidirectional. Changing assumptions would simply require us to
adapt Steps~2.i through~2.iii accordingly.

\section{Multicoloring-based schedules}\label{sec:multic}

Graph $G$ allows us to rephrase the definition of a schedule as follows. We call
a schedule any finite sequence
$\mathcal{S}=\langle S_0,S_1,\ldots,S_{L-1}\rangle$ such that
$S_\ell\subseteq N$ for $0\le\ell\le L-1$, provided
$\bigcup_{\ell=0}^{L-1}S_\ell=N$ and moreover every $S_\ell$ is an independent
set of $G$. The appearance of the notion of an independent set in this
definition leads the way to a special class of schedules, namely those that can
be identified with graph multicolorings \cite{s76}.

For $q\ge 1$, a $q$-coloring of the nodes of $G$ is a mapping from $N$, the
graph's set of nodes, to $\mathbb{N}^q$, where $\mathbb{N}$ is the set of
natural numbers, such that no two of a node's $q$ colors are the same and
besides none of them coincides with any one of any neighbor's $q$ colors. Of
course, the set of nodes receiving one particular color is an independent set.
If $p$ is the total number of colors needed to provide $G$ with a $q$-coloring,
then $N$ is covered by the $p$ independent sets that correspond to colors and
every node is a member of exactly $q$ of these sets. Therefore, letting $L=p$
and identifying each $S_\ell$ with the set of nodes receiving color $\ell$
implies that to every $q$-coloring of the nodes of $G$ there corresponds a
schedule $\mathcal{S}$.

These multicoloring-derived schedules constitute a special case in the sense
that every node of $G$ can be found in exactly the same number of sets ($q$) out
of the $L$ sets that make up the schedule. Clearly, though, there are schedules
that do not correspond to multicolorings. For now we concentrate on those that
do and note that $\mathrm{delivered}(\mathcal{S})\le Pq$ always holds (recall
that $P$ stands for the number of origin-to-destination paths). That is,
the greatest number of packets that the $P$ terminal edges of $D$ can deliver
during the $L$ time slots of schedule $\mathcal{S}$ is $q$ per terminal edge.
These schedules can be further specialized, as follows.

\subsection{Standard coloring}

When $q=1$ every node of $G$ receives exactly one color and
$\mathrm{length}(\mathcal{S})=L\ge\chi(G)$, where $\chi(G)$ is the least number
of colors with which it is possible to provide $G$ with a $1$-coloring, known as
the chromatic number of $G$. Using $T^1(\mathcal{S})$ to denote $T(\mathcal{S})$
in this case, we have
\begin{equation}
T^1(\mathcal{S})\le
\frac{P}{\chi(G)}.
\label{eq:T1}
\end{equation}

\subsection{Standard multicoloring}

Coloring $G$'s nodes optimally in the previous case is minimizing the overall
number of colors. This stems not only from the fact that $q=1$, but more
generally from the fact that $q$ is fixed. We can then generalize and define
$\chi^q(G)$ to be the least number of colors with which it is possible to
provide $G$ with a $q$-coloring. Evidently,
$\chi(G)=\chi^1(G)<\chi^2(G)<\cdots$, so the question of multicoloring $G$'s
nodes optimally when $q$ is not fixed can no longer be viewed as that of
minimizing the overall number of colors needed (as this would readily lead to
$q=1$ and $\chi(G)$ colors). Instead, we look at how efficiently the overall
number of colors is used, i.e., at what the value of $q$ has to be so that
$\chi^q(G)/q$ is minimized. This gives rise to the multichromatic number of $G$,
denoted by $\chi^*(G)$ and given by $\chi^*(G)=\inf_{q\ge 1}\chi^q(G)/q$.
Because this infimum can be shown to be always attained, we use minimum instead
and let $q^*$ be the value of $q$ for which $\chi^*(G)=\chi^{q^*}(G)/q^*$.

Using a $q$-coloring for scheduling amounts to having
$\mathrm{length}(\mathcal{S})=L\ge\chi^q(G)$. In this case, letting
$T^*(\mathcal{S})$ stand for $T(\mathcal{S})$ yields
\begin{equation}
T^*(\mathcal{S})\le
\frac{Pq}{\chi^q(G)}\le
\frac{Pq^*}{\chi^{q^*}(G)}=
\frac{P}{\chi^*(G)}.
\label{eq:T*}
\end{equation}

\subsection{Interleaved multicoloring}

A special class of $q$-colorings is what we call interleaved $q$-colorings
\cite{bg89,b00,yz05}. If $i$ and $j$ are two neighboring nodes of $G$, let
$c^i_1<c^i_2<\cdots<c^i_q$ be the $q$ colors assigned to node $i$ by some
$q$-coloring, and likewise let $c^j_1<c^j_2<\cdots<c^j_q$ be those of node $j$.
We say that this $q$-coloring is interleaved if and only if either
$c^i_1<c^j_1<c^i_2<c^j_2<\cdots<c^i_q<c^j_q$ or
$c^j_1<c^i_1<c^j_2<c^i_2<\cdots<c^j_q<c^i_q$ for all neighbors $i$ and $j$. If
we restrict ourselves to interleaved $q$-colorings, then similarly to what we
did above we use $\chi^q_\mathrm{int}(G)$ to denote the least number of colors
with which it is possible to provide $G$ with an interleaved $q$-coloring, and
similarly define the interleaved multichromatic number of $G$, denoted by
$\chi^*_\mathrm{int}(G)$, to be
$\chi^*_\mathrm{int}(G)=\inf_{q\ge 1}\chi^q_\mathrm{int}(G)/q$. Once again it is
always possible to attain the infimum, so we may take $q^*$ to be the value of
$q$ for which $\chi^*_\mathrm{int}(G)=\chi^{q^*}_\mathrm{int}(G)/q^*$.

As for scheduling based on an interleaved $q$-coloring, it corresponds to having
$\mathrm{length}(\mathcal{S})=L\ge\chi^q_\mathrm{int}(G)$. As before, we use
$T^*_\mathrm{int}(\mathcal{S})$ in lieu of $T(\mathcal{S})$ and obtain
\begin{equation}
T^*_\mathrm{int}(\mathcal{S})\le
\frac{Pq}{\chi^q_\mathrm{int}(G)}\le
\frac{Pq^*}{\chi^{q^*}_\mathrm{int}(G)}=
\frac{P}{\chi^*_\mathrm{int}(G)}.
\label{eq:T*int}
\end{equation}

\subsection{Discussion}

It is a well-known fact that
\begin{equation}
\frac{1}{\chi(G)}\le
\frac{1}{\chi^*_\mathrm{int}(G)}\le
\frac{1}{\chi^*(G)}.
\label{eq:ineq-chi}
\end{equation}
The first inequality follows from the definition of $\chi^*_\mathrm{int}(G)$,
considering that every $1$-coloring is (trivially) interleaved. As for the
second inequality, it follows directly from the definition of $\chi^*(G)$. By
these inequalities, should all of Eqs.~(\ref{eq:T1})--(\ref{eq:T*int}) hold with
equalities, we would have
\begin{equation}
T^1(\mathcal{S})\le
T^*_\mathrm{int}(\mathcal{S})\le
T^*(\mathcal{S}).
\end{equation}
Obtaining equalities in Eqs.~(\ref{eq:T1})--(\ref{eq:T*int}), however, requires
both that $\mathrm{delivered}(\mathcal{S})=Pq$ for $q=1$ or $q=q^*$, as the
case may be, and that $\mathrm{length}(\mathcal{S})=\chi^q(G)$ with the same
possibilities for $q$ or
$\mathrm{length}(\mathcal{S})=\chi^{q^*}_\mathrm{int}(G)$.

While the combined requirements involve the exact solution of NP-hard problems
(finding any of $\chi(G)$, $\chi^*_\mathrm{int}(G)$, and $\chi^*(G)$ is
NP-hard; cf., respectively, \cite{k72}, \cite{bg89}, and \cite{gls81}), the
former one alone is always a property of schedules based on multicolorings when
buffering availability is unbounded. To see that this is so, first recall that
the definition of $\mathrm{delivered}(\mathcal{S})$ refers to a repetition of
the whole schedule as far down in time as needed for any transient effects to
have waned. So, given any of the $P$ source-to-destination paths, we can prove
that $\mathrm{delivered}(\mathcal{S})=Pq$ by arguing inductively about what
happens on such a path during that future repetition of $\mathcal{S}$. The basis
case in this induction is the first directed edge on the path (the one leading
out of the source). The property that every appearance of this edge does indeed
transmit a packet follows trivially from the fact that the source has an endless
supply of new packets to provide whenever needed. Assuming that this also
happens to the next-to-last edge on the path (this is our induction hypothesis)
immediately leads to the same conclusion regarding the last edge, the one on
which $\mathrm{delivered}(\mathcal{S})$ is defined. To see this, let $e$ be the
last edge and $e^-$ the next-to-last one. Because $\mathcal{S}$ is repeated
indefinitely, every time slot $t$ sufficiently far down in time in which $e$
appears is the closing time slot of a window in which both $e$ and $e^-$ appear
exactly $q$ times each. By the induction hypothesis, it follows that at least
one packet is guaranteed to exist for transmission through $e$ at time slot $t$.

Buffering availability, however, is not unbounded, so we must argue for its
finiteness. We do this by recognizing another important property of
multicoloring-based schedules, one that is related to constraints~C1 and~C2
introduced earlier. Because every edge of $D$ (node of $G$) appears the exact
same number $q$ of times in $\mathcal{S}$, there certainly always is a finite
value of $B$, the number of buffer positions per node per path that goes through
it, such that C1 and C2 are satisfied. In all interleaved cases, this value is
$B=1$.

\begin{figure}[t]
\centering
\scalebox{0.75}{\includegraphics{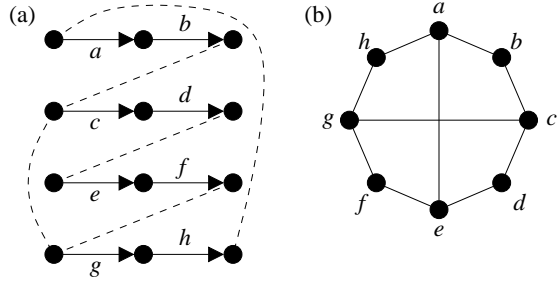}}
\caption{A set of $P=4$ paths (a), with dashed lines indicating all node pairs
representing off-path interference. The resulting graph $G$ is shown in panel
(b). Depending on the schedule $\mathcal{S}$ it is possible to obtain equalities
in all of Eqs.~(\ref{eq:T1})--(\ref{eq:T*int}). The schedules that achieve this
while implying strict inequalities in Eq.~(\ref{eq:ineq-chi}) are:
$\mathcal{S}=\langle\{a,d,g\},\allowbreak
\{b,f,h\},\allowbreak
\{c,e\}\rangle$ for Eq.~(\ref{eq:T1}),
with $T^1(\mathcal{S})=4/3\approx 1.33$;
$\mathcal{S}=\langle\{a,d,f\},\allowbreak
\{b,e,g\},\allowbreak
\{c,f,h\},\allowbreak
\{a,d,g\},\allowbreak
\{b,e,h\},\allowbreak
\{a,c,f\},\allowbreak
\{b,d,g\},\allowbreak
\{c,e,h\}\rangle$
for Eq.~(\ref{eq:T*int}), with $T^*_\mathrm{int}(\mathcal{S})=4/(8/3)=1.5$; and
$\mathcal{S}=\langle\{a,c,f\},\allowbreak
\{b,e,g\},\allowbreak
\{c,e,h\},\allowbreak
\{a,d,g\},\allowbreak
\{b,d,f,h\}\rangle$ 
for Eq.~(\ref{eq:T*}), with $T^*(\mathcal{S})=4/(5/2)=1.6$.}
\label{fig4}
\end{figure}

An example illustrating all of this is presented in Figure~\ref{fig4}, where we
give a set of four source-to-destination paths, the graph $G$ that eventually
results from them, and also the three schedules that result in equalities in
Eqs.~(\ref{eq:T1})--(\ref{eq:T*int}). In this case the two inequalities in
Eq.~(\ref{eq:ineq-chi}) are strict, since it can be shown that
$\chi(G)=3$, $\chi^*_\mathrm{int}(G)=8/3$, and $\chi^*(G)=5/2$ \cite{s76,bg89}. 

\section{Scheduling by edge reversal}\label{sec:ser}

From the three schedules illustrated in Figure~\ref{fig4} it would seem that
finding a schedule $\mathcal{S}$ to maximize $T(\mathcal{S})$ requires that we
give up on the interleaved character of the underlying multicoloring and, along
with it, give up on the equivalent property that edges of $D$ that are
consecutive on some source-to-destination path appear in $\mathcal{S}$
alternately. However, once color interleaving is assumed we are automatically
provided with a principled way to heuristically try and maximize
$T(\mathcal{S})$ by appealing to a curious relationship that exists between
multicolorings and the acyclic orientations of $G$. We now review this heuristic
and later build on it by showing how to adapt it to abandon interleaving only on
occasion during a schedule, aiming at maximizing $T(\mathcal{S})$.

An orientation of $G$ is an assignment of directions to its edges. An
orientation of $G$ is acyclic if no directed cycles are formed. Every acyclic
orientation has a set of sinks (nodes with no edges oriented outward), which by
definition are not neighbors of one another. An acyclic orientation's set of
sinks is then an independent set. The heuristic we now describe, known as
scheduling by edge reversal (SER) \cite{bg89,b96}, is based on the following
property. Should an acyclic orientation be transformed into another by turning
all its sinks into sources (nodes with no edges oriented inward), and should
this be repeated indefinitely, we would obtain an infinite sequence of
independent sets, each given by the set of sinks of the current orientation.
Though infinite, this sequence must necessarily reach a point from which a
certain number of acyclic orientations gets repeated indefinitely. This follows
from the facts that there are only finitely many acyclic orientations of $G$ and
that turning one of them into the next is a deterministic process.

The orientations that participate in this cyclic repetition, henceforth called
a period, have the property that every node of $G$ appears as a sink in the same
number of orientations. Furthermore, any two neighboring nodes of $G$ are sinks
in alternating orientations, regardless of whether the period has already been
reached or not. It clearly follows that the sets of sinks in a period constitute
a schedule that is based on an interleaved multicoloring. Depending on the very
first acyclic orientation in the operation of SER more than one period can
eventually be reached. The different periods' properties vary from one to
another, but it can be shown that at least one of them corresponds to the
optimal interleaved multicoloring, i.e., the one that yields
$\chi^*_\mathrm{int}(G)$ \cite{bg89}. The heuristic nature of SER is then
revealed by the need to determine an appropriate initial acyclic orientation.

Determining a schedule $\mathcal{S}$ by SER follows the algorithm given next. We
use $\omega_0,\omega_1,\ldots$ to denote the sequence of acyclic orientations of
$G$. For $t=0,1,\ldots$, we denote by $\mathrm{Sinks}(\omega_t)$ the set of
sinks of $\omega_t$.

\bigskip\noindent
\textsc{Algorithm} SER:
\begin{enumerate}
\item Choose $\omega_0$.
\item $t := 0$.
\item Obtain $\omega_{t+1}$ from $\omega_t$.
\item If the period has not yet occurred, then $t:=t+1$; go to Step 3. If it
has, then let $p(\omega_0)$ be its number of orientations, $m(\omega_0)$ the
number of times any node appears in them as a sink, and
$\omega_k,\omega_{k+1},\ldots,\omega_{k+p(\omega_0)-1}$ the orientations
themselves. Output
$$
\mathcal{S}=
\langle\mathrm{Sinks}(\omega_k),\mathrm{Sinks}(\omega_{k+1}),\ldots,
\mathrm{Sinks}(\omega_{k+p(\omega_0)-1})\rangle
$$
and
$$
T(\mathcal{S})=
\frac{Pm(\omega_0)}{p(\omega_0)}.
$$
\end{enumerate}

In this algorithm, the explicit dependency of both $p(\omega_0)$ and
$m(\omega_0)$ on $\omega_0$ is meant to emphasize that, implicitly, the two
quantities are already determined when in Step~1 the initial orientation
$\omega_0$ is chosen. As with the very existence of the period, this follows
from the fact that the algorithm's Step~3 is deterministic, so there really is
no choice regarding the period to be reached once $\omega_0$ has been fixed. The
role played by the two quantities is precisely to characterize the interleaved
multicoloring mentioned above. That is, the period reached from $\omega_0$ can
be regarded as assigning $q=m(\omega_0)$ distinct colors to each node of $G$
using a total number $p=p(\omega_0)$ of colors. Equivalently, it can be
regarded as a schedule $\mathcal{S}$ for which
$\mathrm{delivered}(\mathcal{S})=Pq=Pm(\omega_0)$ (where the first equality is
true of all multicoloring-based schedules, as we discussed in
Section~\ref{sec:multic}) and $\mathrm{length}(\mathcal{S})=L=p=p(\omega_0)$. By
Eq.~(\ref{eq:throughput}), the final determination of $T(\mathcal{S})$ follows
easily.

As a final observation, we note that, although the knowledge of $p(\omega_0)$
and $m(\omega_0)$ after Step~1 is only implicit, it can be shown that the ratio
$m(\omega_0)/p(\omega_0)$ can be known explicitly at that point \cite{bg89}. It
might then seem that the remainder of the algorithm is useless, since the value
of $T(\mathcal{S})$ can be calculated right after Step~1. But the reason why the
remaining steps are needed, of course, is that $\mathcal{S}$ itself needs to be
found, not just the $T(\mathcal{S})$ that quantifies its performance.

\section{Improving on SER}\label{sec:sera}

\begin{figure}[t]
\centering
\scalebox{0.75}{\includegraphics{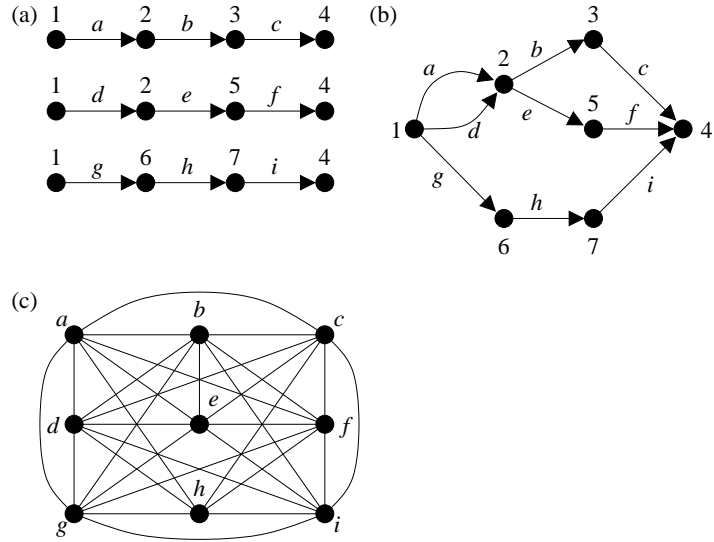}}
\caption{A set of $P=3$ directed paths (a), the resulting directed multigraph
$D$ (b), and the resulting undirected graph $G$ (c). The optimal SER schedule is
$\mathcal{S}=\langle\{a\},\allowbreak
\{b\},\allowbreak
\{c,g\},\allowbreak
\{d,i\},\allowbreak
\{e,h\},\allowbreak
\{f\}\rangle$,
yielding $T(\mathcal{S})=3/6=0.5$. An alternative schedule that does not comply
with the SER alternance condition, with $B=2$, is
$\mathcal{S}=\langle\{g,c\},\allowbreak
\{g,f\},\allowbreak
\{h,b\},\allowbreak
\{h,e\},\allowbreak
\{i,a\},\allowbreak
\{i,d\}\rangle$,
which results in an improvement to $T(\mathcal{S})=4/6\approx 0.67$.}
\label{fig5}
\end{figure}

In Figure~\ref{fig5} we provide an example to illustrate why giving up
interleaving may yield a schedule $\mathcal{S}$ of higher $T(\mathcal{S})$. The
general idea is that, given $B$, it may be possible to schedule a given
transmission sooner than it normally would be scheduled by SER, provided there
is a packet to be transmitted in the buffers of the sending node in $D$ and the
receiving node has an available buffer position for the path in question. While
under SER two transmissions sharing a buffer alternate with each other in any
schedule (and then $B=1$ always suffices), disrupting this alternance implies
that all buffering is to be managed in detail.

In the example of Figure~\ref{fig5} transmissions $g$, $h$, and $i$ are
scheduled without regard to alternance if $B=2$. While this results in improved
performance (more packets delivered to node $4$ per time slot), it is important
to realize that this is in great part made possible by the structure of $D$ in
this example. Even though all three paths lead from node $1$ to node $4$, two of
them are poised to interfere with each other particularly heavily by virtue of
sharing node $2$. The consequence of this is that transmissions on these paths
will occur less in parallel than they might otherwise. But since $B=2$ buffering
positions are available per node per path, the path that goes through nodes $6$
and $7$ can compensate for this by transmitting twice as much traffic (thence
the double occurrence of $g$ in a row, and also of $h$ and $i$, for each
repetition of the schedule). This, however, is never detrimental to the traffic
on the other two paths: all that is being done is to seize the opportunity to
transmit in time slots that would otherwise go unused.

Implementing the careful buffer management alluded to above requires that we
look at the dynamics of acyclic-orientation transformation under SER in more
detail. Given any acyclic orientation $\omega$ of $G$, the node set $N$ of $G$
can be partitioned into independent sets $I_1,I_2,\ldots,I_d$ such that $I_1$ is
the set of all sinks in $G$ according to $\omega$, $I_2$ is the set of all sinks
we would obtain if all nodes in $I_1$ were to be eliminated, and so on. In this
partition, known as a sink decomposition, $d$ is the number of nodes on a
longest directed path of $G$ according to $\omega$. When $\omega$ is turned into
$\omega'$ by SER a new sink decomposition is obtained, call it
$I'_1,I'_2,\ldots,I'_{d'}$, such that $I'_1=I_2$, $I'_2\supseteq I_3$, etc.,
with $d'\le d$. The reason why equality need not hold in all cases, but set
containment instead, is that each $I_k$ may get enlarged by some of the previous
orientation's sinks before becoming $I'_{k-1}$.

\begin{figure}[t]
\centering
\scalebox{0.75}{\includegraphics{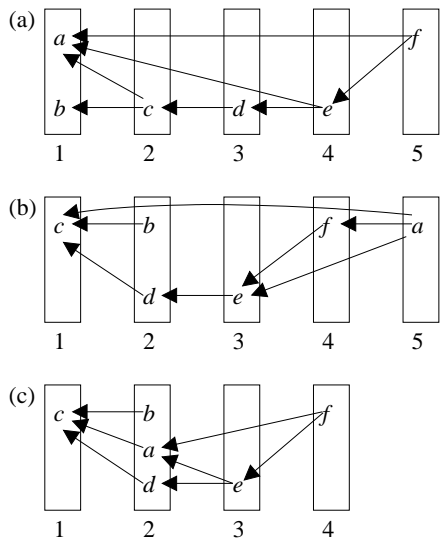}}
\caption{
Each set in a sink decomposition is represented by a rectangular box and
numbered to indicate the set's subscript. Note that directed edges refer to
acyclic orientations of $G$. Applying SER to the sink decomposition in panel (a)
yields the one in panel (b). In this transformation both $a$ and $b$ are turned
into sources. The alternative of using SERA, on the other hand, makes it
possible for $a$ to be placed in a lower-subscript set, avoiding the
transformation into a source and yielding the sink decomposition in panel (c).
This can be done only because the set to which $a$ is added contains none of its
neighbors in $G$. Assuming that transmissions $a$, $e$, and $f$ are initially
arranged consecutively in one of the $P$ paths in the order $e,a,f$, we have
$i^-=e$, $i=a$, and $i^+=f$. We also have, in reference to panel (a), $k^-=4$,
$k=2$, and $k^+=5$. Then the additional conditions for the move from (a) to (c)
to occur are that the buffers shared by transmissions $e$ and $a$ contain at
least one packet (since $k<k^-$), and that those shared by $a$ and $f$ have room
for at least one packet (since $k<k^+$).}
\label{fig6}
\end{figure}

It is then possible to regard the operation of SER as simply a recipe for
manipulating sink decompositions. At each iteration the set containing the
current sinks is eliminated and its nodes are redistributed through the other
sets. The remaining sets are renumbered by decrementing their subscripts by $1$
and a new, greatest-subscript set may have to be created. The rule for
redistributing each of the former sinks is to find the set of greatest subscript
containing one of the node's neighbors in $G$, say $I_k$, and then place the
node in set $I_{k+1}$. This is illustrated in panels (a) and (b) of
Figure~\ref{fig6}.

Altering this rule is the core of our modified SER, henceforth called SER with
advancement (SERA). If $i$ is the sink in question, the operation of SERA is
based on placing node $i$ in the least-subscript set that does not contain a
neighbor of $i$ in $G$. This clearly maintains acyclicity just as the previous
rule does, but now the former sink is not necessarily turned into a source, but
rather into a node that can have edges oriented both inward and outward by the
current orientation, respectively from nodes in sets of greater subscripts and
to nodes in sets of lesser subscripts. Additionally, while this alternative
placement of node $i$ does favor it by virtue of lowering the number of time
slots that need to go by before it is a sink once again, clearly there is no
detriment to any of the other transmissions, which will assuredly become sinks
no later than they would otherwise.

As we mentioned, however, unlike SER this rule can only be applied as a function
of $B$ and the buffering-related constraints we mentioned. Suppose that $i$ is
preceded by transmission $i^-$ and succeeded by transmission $i^+$ on the
original path out of $\mathcal{P}_1,\mathcal{P}_2,\ldots,\mathcal{P}_P$ to which
it belongs. Of course, both $i^-$ and $i^+$ are nodes of $G$ as well. Suppose
further that these two nodes are in sets $I_{k^-}$ and $I_{k^+}$, respectively,
and that we are attempting to place node $i$ in set $I_k$. The further
constraints to be satisfied are the following. If $k<k^-$, then the buffers
shared by transmissions $i^-$ and $i$ must contain at least one packet to be
transmitted. If $k<k^+$, then the buffers shared by transmissions $i$ and $i^+$
must contain room to store at least one more packet. This can all be implemented
rather easily by keeping a dynamic record of all buffers. A simple case of
evolving sink decompositions in the SERA style is shown in panels (a) and (c) of
Figure~\ref{fig6}.

SERA, like SER, operates on finitely many possibilities and deterministically.
A ``possibility'' is no longer simply an acyclic orientation, but instead an
acyclic orientation together with a configuration of buffer occupation. In any
event, periodic behavior is still guaranteed to occur and we go on denoting by
$p(\omega_0)$ the number of possibilities in the period that one reaches from
$\omega_0$. The notion behind $m(\omega_0)$, however, has been lost together
with the certainty of interleaving, since SERA does not guarantee that every
node of $G$ is a sink in the period the same number of times. For $i\in N$, an
alternative definition is that of $m_i(\omega_0)$, which we henceforth use to
denote the number of times node $i$ is a sink in the period, not necessarily the
same for all nodes.

Determining the schedule $\mathcal{S}$ through SERA proceeds according to the
following algorithm.

\bigskip\noindent
\textsc{Algorithm} SERA:
\begin{enumerate}
\item Choose $\omega_0$.
\item $t := 0$.
\item Obtain $\omega_{t+1}$ from $\omega_t$, employing advancement as described.
\item If the period has not yet occurred, then $t:=t+1$; go to Step 3. If it
has, then let $p(\omega_0)$ be its number of orientations (with associated
buffer-occupation configurations), $m_i(\omega_0)$ the number of times node $i$
appears in them as a sink, and
$\omega_k,\omega_{k+1},\ldots,\omega_{k+p(\omega_0)-1}$ the orientations
themselves. Output
$$
\mathcal{S}=
\langle\mathrm{Sinks}(\omega_k),\mathrm{Sinks}(\omega_{k+1}),\ldots,
\mathrm{Sinks}(\omega_{k+p(\omega_0)-1})\rangle
$$
and
$$
T(\mathcal{S})=
\frac{\sum_{i\in T}m_i(\omega_0)}{p(\omega_0)},
$$
where $T$ is the set of the nodes of $G$ that correspond to terminal edges of
the paths $\mathcal{P}_1,\mathcal{P}_2,\ldots,\mathcal{P}_P$.
\end{enumerate}

In this algorithm, note that the determination of $T(\mathcal{S})$ generalizes
what is done in SER. This is achieved by adopting
$\mathrm{delivered}(\mathcal{S})=\sum_{i\in T}m_i(\omega_0)$ while maintaining
$\mathrm{length}(\mathcal{S})=p(\omega_0)$ in Eq.~(\ref{eq:throughput}).
Particularizing this to the case of SER yields
$\mathrm{delivered}(\mathcal{S})=Pm(\omega_0)$, as desired, since
$m_i(\omega_0)$ becomes $m(\omega_0)$ for any node $i$ of $G$ and moreover
$\vert T\vert=P$.

\section{Methods}\label{sec:methods}

We have conducted extensive computational experiments to evaluate SER and SERA,
the latter with a few different values for the buffering parameter $B$. Before
we present results in Section~\ref{sec:results}, here we pause to introduce the
methodology that was followed. This includes selecting the network topology that
eventually leads to graph $G$ and the choice of the initial acyclic orientation
of $G$.

\subsection{Topology generation}

We generated $1600$ networks by placing $n$ nodes inside a square of side
$1500$. For each network the first node was positioned at the square's center.
Given the nodes' communication (or interference) radius $R$, and with it the
neighborhood relation among nodes (i.e., two nodes are neighbors of each other
if and only if the Euclidean distance between them is no greater than $R$), we
proceeded to positioning the remaining nodes randomly, one at a time.
Positioning a node was subject to the constraints that it would have at least
one neighbor, that no node would have more than $\Delta$ neighbors, and moreover
that no two nodes would be closer to each other than $25$ units of Euclidean
distance. Repeated attempts at positioning nodes while satisfying these
constraints were not allowed to number more than $1000$ per network. When this
limit was reached the growing network was wiped clean and a new one was started.
The value of $R$ was determined so that, had the nodes been positioned uniformly
at random, a randomly chosen radius-$R$ circle would have expected density
proportional to $\Delta/R^2$ and about the same density as the whole network,
i.e., $\Delta/R^2\propto n$. Choosing the proportionality constant to yield
$R=200$ for $n=80$ and $\Delta=4$ results in the formula
$R=200\sqrt{20\Delta/n}$. Of the $1600$ networks thus generated, there are $100$
networks for each combination of $n\in\{60,80,100,120\}$ and
$\Delta\in\{4,8,16,32\}$.

For each network we generated $50n$ sets of paths
$\mathcal{P}_1,\mathcal{P}_2,\ldots,\mathcal{P}_P$, each $100$ sets
corresponding to a different value of $P$. Each of the sets resulted in a
different $D$, then in $G$, as explained in Sections~\ref{sec:form}
and~\ref{sec:transf}. The $50n$ sets comprise $100$ groups of $n/2$ sets each.
The first of these sets for a group has $P=1$ and the single path it contains is
the shortest path from a randomly chosen node to another in the network. Each
new set in the group is the previous one enlarged by the addition of a new path,
obtained by selecting two distinct nodes randomly, provided they do not already
participate in the previous set. This goes on until $P=n/2$, so in the last set
every one of the $n$ nodes participates as either the origin or the destination
of one of the $P$ paths. For the sake of normalization, the results we present
for $T(\mathcal{S})$, given for $P=1,2,\ldots,n/2$, are shown against the ratio
$P'=2P/n\in(0,1]$.

\subsection{Initial acyclic orientation}

Once $G$ has been built for a fixed network and a fixed set of paths, the
acyclic orientation $\omega_0$ of $G$ has to be determined. Our general approach
is to label every node of $G$ with a different number and then to direct each
edge from the node that has the higher number to the one that has the lower.
Although the resulting orientation is clearly acyclic, we are left with the
problem of labeling the nodes. We approach this problem by resorting to the
paths $\mathcal{P}_1,\mathcal{P}_2,\ldots,\mathcal{P}_P$ from which $G$
resulted, since the nodes of $G$ are in one-to-one correspondence with the
directed edges on the paths. It then suffices to number the paths' edges.

We consider four numbering schemes:
\begin{list}{}{}
\item[ND-BF.] The paths are organized in the nondecreasing order of their
numbers of edges (ties are broken by increasing path number). The edges are then
numbered breadth-first from the path's origins, given this organization of the
paths.
\item[ND-DF.] The paths are organized in the nondecreasing order of their
numbers of edges (ties are broken by increasing path number). The edges are then
numbered depth-first from the paths' origins, given this organization of the
paths.
\item[NI-BF.] The paths are organized in the nonincreasing order of their
numbers of edges (ties are broken by increasing path number). The edges are then
numbered breadth-first from the paths' origins, given this organization of the
paths.
\item[NI-DF.] The paths are organized in the nonincreasing order of their
numbers of edges (ties are broken by increasing path number). The edges are then
numbered depth-first from the paths' origins, given this organization of the
paths.
\end{list}

\section{Computational results}\label{sec:results}

We divide our results into two categories. First we give statistics on the
$1600$ networks generated for evaluation of the algorithms. Then we report on
the values obtained for $T(\mathcal{S})$ by SER and SERA.

One of the statistics is particularly useful: despite its simplicity, we have
found it to correlate with the SERA results in a fairly direct way. This
statistic is based on a function of $G$ that aims to quantify how the
interference among the initial $P$ directed paths is reflected in the structure
of $G$. We denote this function by $\rho(G)$ and let it be such that
\begin{equation}
\rho(G)=\frac{P\vert E'\vert}{\sum_{p=1}^P\vert Y_p\vert}.
\end{equation}
In this equation, recall that the sets $Y_1,Y_2,\ldots,Y_P$, one for each of the
initial directed paths, contain the edges that ultimately become the nodes of
$G$. Thus, $\sum_{p=1}^P\vert Y_p\vert/P$ is the average number of edges on a
path. Moreover, we let $E'\subseteq E$ be the set of $G$'s edges whose end nodes
correspond to edges of distinct paths. In words, then, $\rho(G)$ is the average
number of off-path transmissions that interfere with the transmissions of a path
having the average number of edges.

\subsection{Properties of the networks generated}

\begin{figure}[p]
\centering
\scalebox{0.55}{\includegraphics{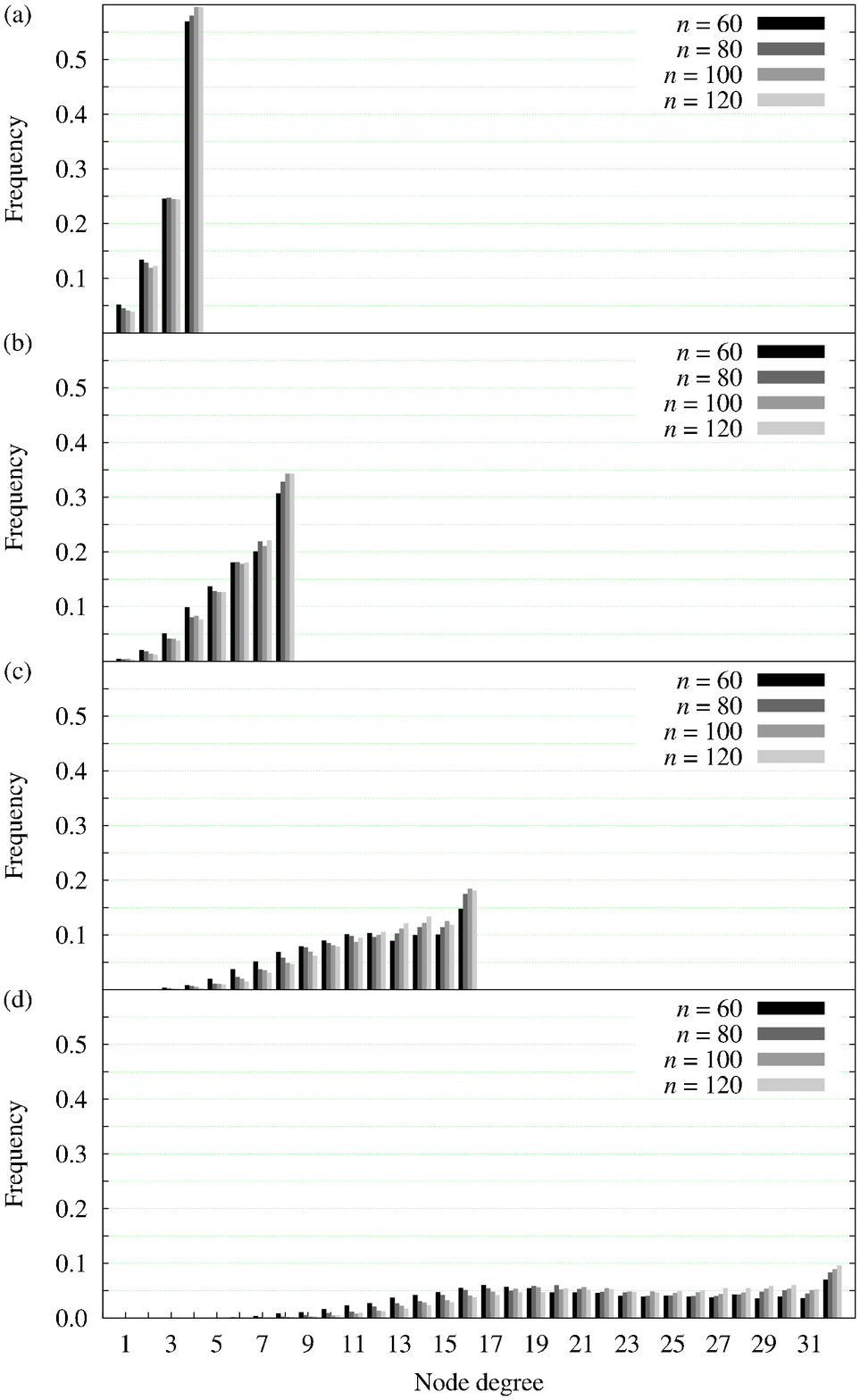}}
\caption{Degree distributions of the $1600$ networks, for $\Delta=4$ (a),
$\Delta=8$ (b), $\Delta=16$ (c), and $\Delta=32$ (d). For each combination of
$n$ and $\Delta$ the distribution refers to $100$ networks.}
\label{fig7}
\end{figure}

\begin{figure}[p]
\centering
\scalebox{0.55}{\includegraphics{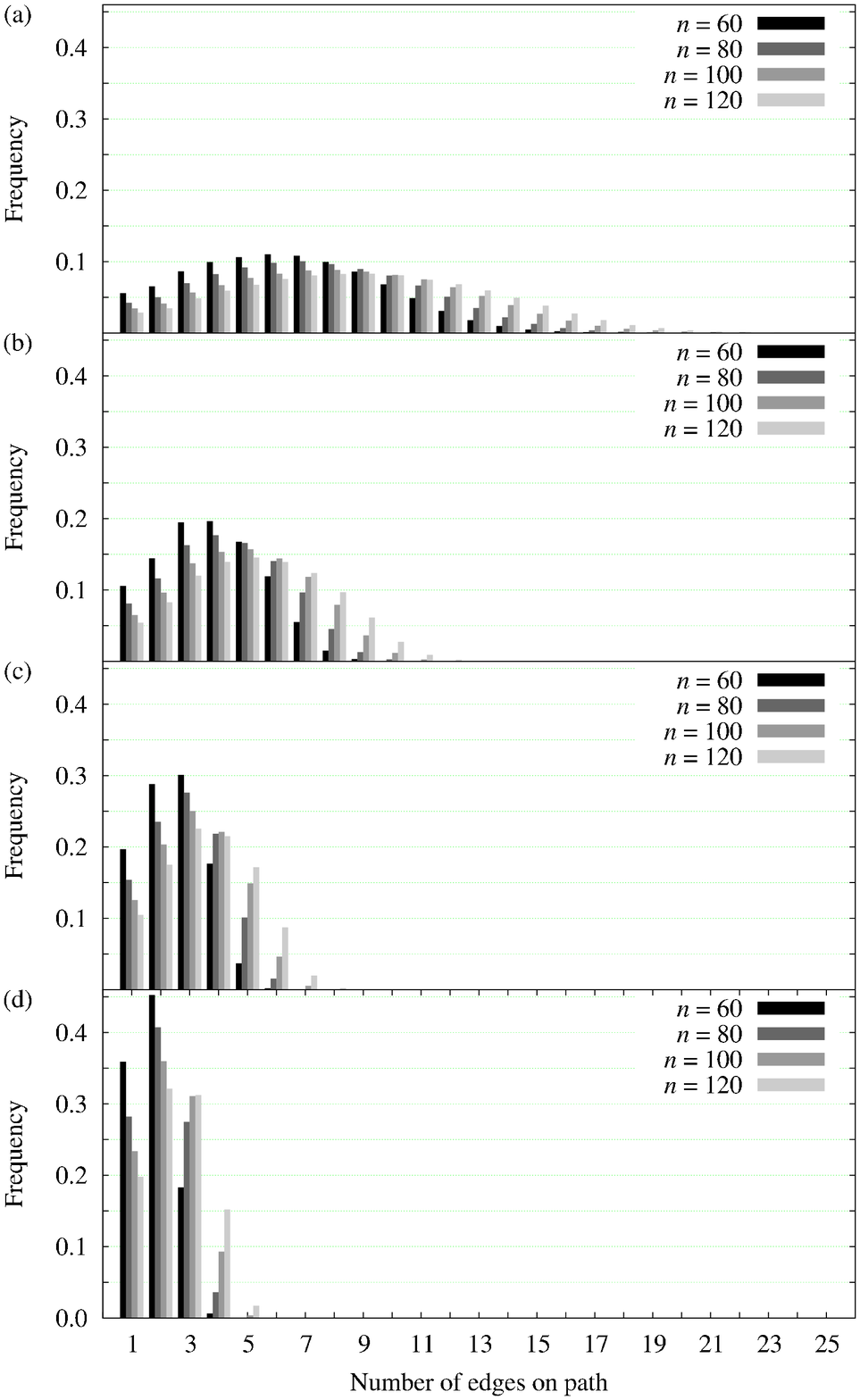}}
\caption{Path-size (number of edges) distributions of the $1600$ networks, for
$\Delta=4$ (a), $\Delta=8$ (b), $\Delta=16$ (c), and $\Delta=32$ (d). For each
combination of $n$ and $\Delta$ the distribution refers to $100$ networks and to
$100$ path sets for each network, each set comprising $P=n/2$ paths.}
\label{fig8}
\end{figure}

The $1600$ networks' distributions of node degrees are given in
Figure~\ref{fig7}, which contains one panel for each of the four values of
$\Delta$ and all four values of $n$. Their distributions of the numbers of edges
on the $P$ paths for $P=n/2$ are given in Figure~\ref{fig8}, again with one
panel for each of the four values of $\Delta$ and all four values of $n$.

We see from Figure~\ref{fig7} that the degree distributions peak at the degree
$\Delta$, falling approximately linearly toward the lower degrees (except for
$\Delta=32$, where a plateau is observed midway). Furthermore, the lowest
observed degree grows with $\Delta$, which is expected from the formula that
gives the radius $R$ as an increasing function of $\Delta$. We also see from the
figure that these distributions are approximately independent of the value of
$n$ for fixed $\Delta$; in reference to that same formula, we see that letting
$R$ decrease with $n$ does indeed have the expected effect of maintaining an
approximately uniform node density throughout the containing square.

We also expect path sizes to be smaller as $\Delta$ increases, and this is in
fact what Figure~\ref{fig8} shows. In fact, larger $\Delta$ values decreases
the variability of path sizes, which moreover get concentrated around an ever
smaller mean. For fixed $\Delta$, what we see in the figure is a consistent
shift to the right (i.e., greater path sizes) as $n$ grows. This reflects the
fact that larger $n$ for fixed $\Delta$ leads to smaller $R$, thus to longer
paths.

These observations are summarized in Table~\ref{tab1}, where the mean degree
and mean path size are given for each combination of $n$ and $\Delta$ values.
This table also shows the average value of $\rho(G)$, defined above as an
indicator of how much interference there is in $G$ among all $P$ paths, when $G$
refers to $P=n/2$. For fixed $n$, it is curious to observe that $\rho(G)$
decreases as $\Delta$ is decreased from $32$ through $8$, but then appears to
flatten out or even rebound slightly as $\Delta$ is further decreased to $4$.
Each of these averages corresponds to $10^4$ $G$ instances ($100$ instances
corresponding to the $P=n/2$ case of each of the $100$ networks for fixed $n$
and $\Delta$) and is significant to the extent of the confidence interval
reported for it in the table's rightmost column. As we demonstrate shortly, the
peculiar behavior of $\rho(G)$ helps explain a lot of what is observed with
respect to how $T(\mathcal{S})$ behaves in the case of SERA.

\begin{table}[t]
\centering
\caption{Mean values of the distributions in Figures~\ref{fig7} and~\ref{fig8},
and the average $\rho(G)$ values for the $10^4$ $G$ instances corresponding to
each combination of $n$ and $\Delta$ when $P=n/2$. Confidence intervals refer to
these averages and are given at the $95\%$ level.}
\begin{tabular}{cccccc}
\hline
$n$ & $\Delta$ & Mean   & Mean      & \multicolumn{2}{c}{ $\rho(G)$} \\ \cline{5-6}
    &          & degree & path size & Average & Conf.\ int. \\ \hline
60  & 4        & 3.33   & 7.46      & 0.4     & 0.06 \\
    & 8        & 6.22   & 4.85      & 0.37    & 0.06 \\
    & 16       & 11.67  & 3.57      & 0.4     & 0.05 \\
    & 32       & 21.23  & 2.84      & 0.58    & 0.05 \\ \hline
80  & 4        & 3.36   & 8.32      & 0.37    & 0.06 \\
    & 8        & 6.37   & 5.36      & 0.35    & 0.06 \\
    & 16       & 12.17  & 3.92      & 0.39    & 0.05 \\
    & 32       & 22.36  & 3.06      & 0.59    & 0.05 \\ \hline
100 & 4        & 3.40   & 9.3       & 0.36    & 0.06 \\
    & 8        & 6.40   & 5.86      & 0.34    & 0.07 \\
    & 16       & 12.40  & 4.22      & 0.38    & 0.06 \\
    & 32       & 23.09  & 3.27      & 0.6     & 0.05 \\ \hline
120 & 4        & 3.40   & 9.95      & 0.34    & 0.06 \\
    & 8        & 6.45   & 6.28      & 0.33    & 0.07 \\
    & 16       & 12.50  & 4.52      & 0.38    & 0.06 \\
    & 32       & 23.59  & 3.47      & 0.58    & 0.05 \\ \hline
\end{tabular}

\label{tab1}
\end{table}

\subsection{Results}

\begin{figure}[p]
\centering
\scalebox{0.55}{\includegraphics{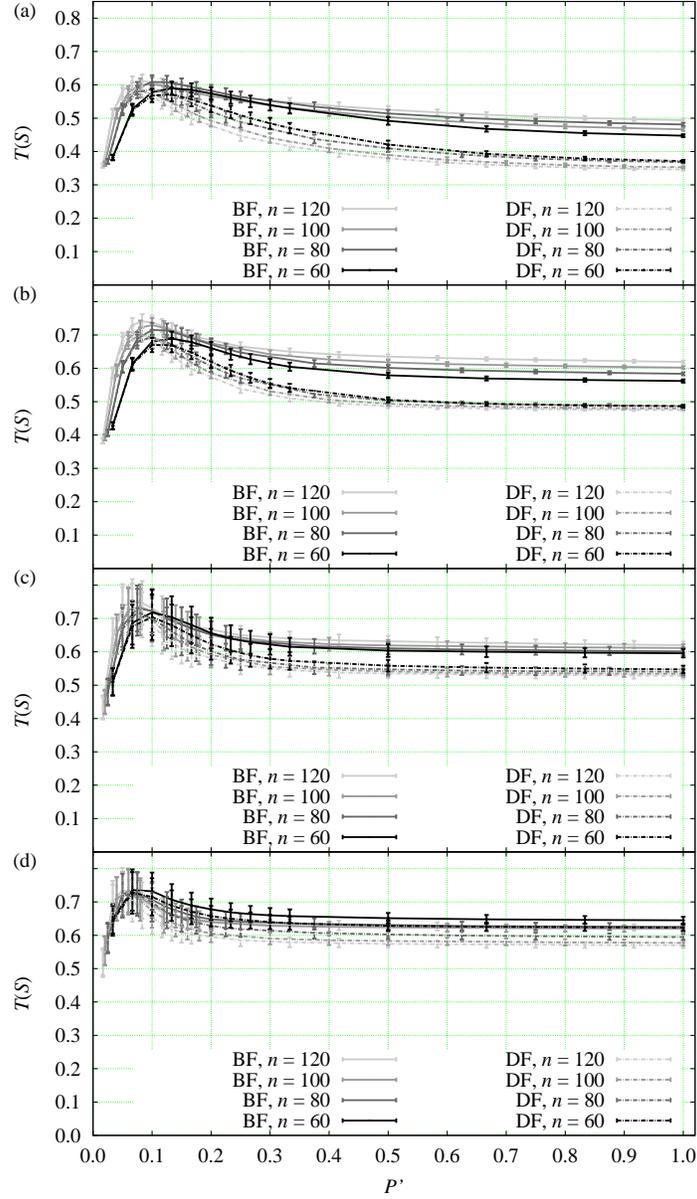}}
\caption{Behavior of $T(\mathcal{S})$ for SER under the two numbering schemes
ND-BF and ND-DF, with $\Delta=4$ (a), $\Delta=8$ (b), $\Delta=16$ (c), and
$\Delta=32$ (d). Data are averages over the $10^4$ $G$ instances that correspond
to each combination of $n$ and $\Delta$ for each value of $P$. Error bars are
based on confidence intervals at the $95\%$ level.}
\label{fig9}
\end{figure}

Our results for SER are given in Figure~\ref{fig9} as plots of $T(\mathcal{S})$
against the $P'$ ratio introduced in Section~\ref{sec:methods}. Each of the
figure's four panels is specific to a fixed $\Delta$ value and shows a plot for
each value of $n$ combined with either the ND-BF or the ND-DF numbering scheme.
All results relating to the NI-BF and NI-DF schemes are omitted, as we found
them to be statistically indistinguishable from their ND counterparts. From this
figure it seems clear that, as $P'$ increases (i.e., as the number of paths $P$
grows towards $n/2$), the superiority of the BF schemes over the DF schemes
becomes apparent, more pronouncedly so for the lower values of $\Delta$. The
reason why the BF schemes tend to perform better than the DF schemes should be
intuitively clear: the BF schemes number the transmissions that are closer to
the paths' origins first, therefore with the lowest numbers. As the initial
acyclic orientation of $G$ is built from these numbers, the first sinks during
the operation of SER will correspond to starting parallel traffic on as many
paths as possible. Overall it also seems that larger values of $n$ lead to
better performance for fixed $\Delta$, but the distinction appears to be only
marginal and is sometimes obscured by the confidence intervals.

\begin{figure}[p]
\centering
\scalebox{0.55}{\includegraphics{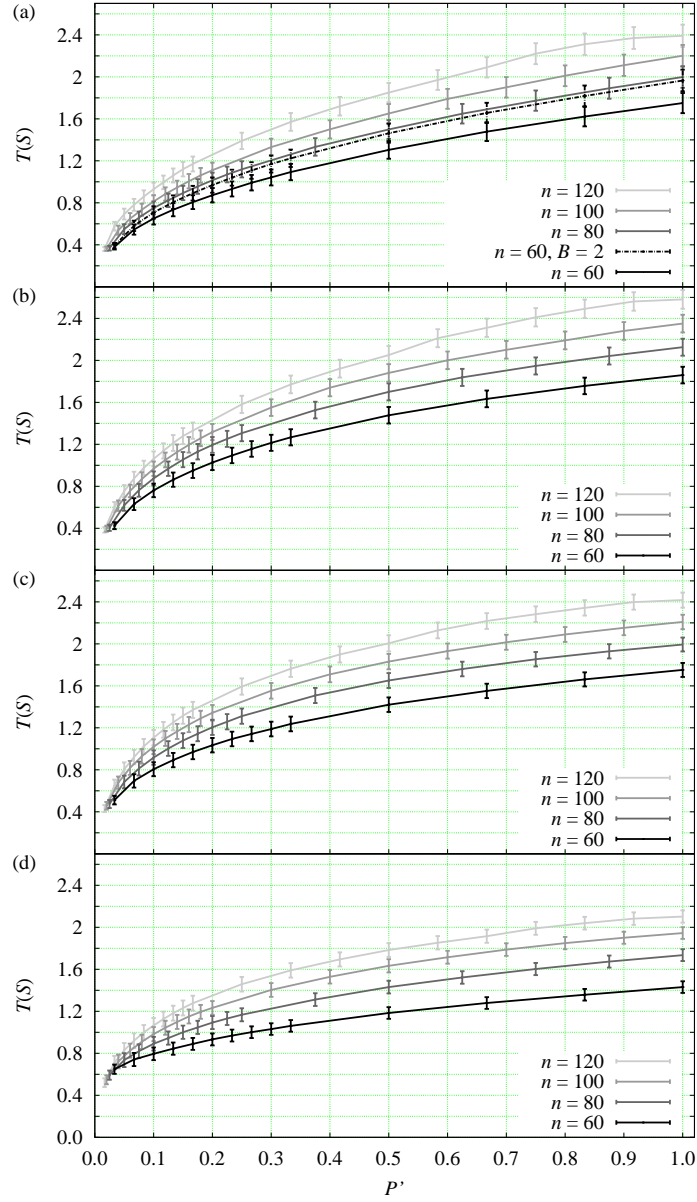}}
\caption{Behavior of $T(\mathcal{S})$ for SERA under the numbering scheme
ND-BF, with $\Delta=4$ (a), $\Delta=8$ (b), $\Delta=16$ (c), and $\Delta=32$
(d). Data are averages over the $10^4$ $G$ instances that correspond to each
combination of $n$ and $\Delta$ for each value of $P$. Error bars are based on
confidence intervals at the $95\%$ level.}
\label{fig10}
\end{figure}

A similar set of plots is given in Figure~\ref{fig10}, now displaying our
results for SERA as plots of $T(\mathcal{S})$ against the ratio $P'$. Once
again there is one panel for each value of $\Delta$, and once again several
possibilities regarding the numbering schemes are omitted because of statistical
indistinguishability. This is also true of the various possibilities for the
value of $B$, with the single exception we mention shortly. Thus, most plots
correspond to the ND-BF numbering scheme with $B=1$. The single exception is
that of $n=60$ with $\Delta=4$, for which we also report on the $B=2$ case. For
fixed $\Delta$ and $P'$, increasing $n$ also leads to increased
$T(\mathcal{S})$. In the particular case of $n=60$ and $\Delta=4$, increasing
$B$ from $1$ to $2$ also causes $T(\mathcal{S})$ to increase.

As we fix $\Delta$, $n$, and $P'$, Figures~\ref{fig9} and~\ref{fig10} reveal
that $T(\mathcal{S})$ is consistently higher for SERA than it is for SER (by a
factor of about $2$ to $4$) across all values for these quantities, thereby
establishing the superiority of the former algorithm over the latter. For
sufficiently large $n$ this occurs for the same value of $B$ (that is, for
$B=1$), which furthermore establishes that this superiority does not in general
depend on the availability of more buffering space. It is, instead, determined
solely by the elimination in SERA of the mandatory alternance of interfering
transmissions in SER.

Fixing $n$ and $P'$ while varying $\Delta$ (i.e., moving between panels) yields
further interesting insight about the two algorithms. While for SER increasing
$\Delta$ under these conditions causes $T(\mathcal{S})$ to increase
monotonically (though sometimes almost imperceptibly) for the same numbering
scheme, doing the same for SERA for constant $B$ leads $T(\mathcal{S})$ to
behave in a markedly non-monotonic way. In fact, as $\Delta$ is increased from
$4$ to $8$ there is also an increase in $T(\mathcal{S})$, but increasing
$\Delta$ further through $\Delta=32$ leads to decreases in $T(\mathcal{S})$.
As we anticipated earlier, this is fully analogous to the behavior of $\rho(G)$
as $\Delta$ is increased in the same way while all else remains constant. This
suggests that what determines the relative behavior of $T(\mathcal{S})$ in
these circumstances is the intensity of inter-path interference as it gets
shaped by the structure of $G$. In other words, $T(\mathcal{S})$ and $\rho(G)$
tend to vary along somewhat inverse trends with respect to each other.

\section{Discussion}\label{sec:disc}

When SER is used, it follows from our discussion in Sections~\ref{sec:multic}
and~\ref{sec:ser} that $T(\mathcal{S})=T^*_\mathrm{int}(\mathcal{S})$. By
Eq.~(\ref{eq:T*int}), we then have
\begin{equation}
T(\mathcal{S})\le\frac{P}{\chi^*_\mathrm{int}(G)},
\end{equation}
where achieving equality requires that we choose $\omega_0$ optimally. Now let
$\varphi(G)=\max\{\omega(G),\vert N\vert/\alpha(G)\}$, where $\omega(G)$ is the
number of nodes in the largest clique of $G$ and $\alpha(G)$ is the number of
nodes in the largest independent set of $G$. It can be shown that
$\chi^*_\mathrm{int}(G)\ge\varphi(G)$,\footnote{See \cite{l03}, where the
interleaved multichromatic number of $G$ is referred to as $G$'s circular
chromatic number, and references therein.} whence
\begin{equation}
T(\mathcal{S})\le\frac{P}{\varphi(G)}=\frac{P'n}{2\varphi(G)},
\label{eq:bound}
\end{equation}
where we have taken into account the way we handle $P$ in all our experiments.
We see then that $T(\mathcal{S})$ is bounded from above by the fraction of
$n/2$ given by $P'/\varphi(G)$. For fixed $P'$, this fraction tends to be
small if the largest clique of $G$ is large or its largest independent set is
small, whichever is more influential on $\varphi(G)$. Either possibility
bespeaks the presence of considerable interference among the transmissions
represented by the $\vert N\vert$ nodes of $G$.

Of course, in general we have no practical way of knowing how close each
$\omega_0$ we choose is to being the optimal starting point for SER, nor of
knowing how different $\chi^*_\mathrm{int}(G)$ and $\varphi(G)$ are for the $G$
instances we use. So the bound given in Eq.~(\ref{eq:bound}), located somewhere
between $30P'/\varphi(G)$ and $60P'/\varphi(G)$ for our values of $n$,
cannot be used as a guide to assessing how low the $T(\mathcal{S})$ values
shown in Figure~\ref{fig9} really are. But the bound's sensitivity to growing
interference in $G$ does provide some guidance, since all plots in the figure
become flat from about $P'=0.3$, regardless of the value of $n$ or the numbering
scheme used. Perhaps every $G$ corresponding to such values of $P'$ share some
structural property, like a very large clique or only very small independent
sets, that renders the resulting values of $T(\mathcal{S})$ oblivious to all
else.

As for SERA, since the schedules it produces depart from a strict
characterization as multicolorings of $G$, no upper bounds on $T(\mathcal{S})$
are known to us. Nevertheless, a comparison with SER as provided by
Figures~\ref{fig9} and~\ref{fig10} reveals that $T(\mathcal{S})$ for SERA
surpasses $T(\mathcal{S})$ for SER by a substantial margin, and also that SERA
is capable of finding ways to improve $T(\mathcal{S})$ somewhat even as $P'$
grows. If our observation above regarding the structure of $G$ as an inherent
barrier to improving $T(\mathcal{S})$ as $P'$ grows is true, then the barrier's
effects under SERA are considerably attenuated. This, we believe, is to be
attributed to SERA's aggressively opportunistic approach of abandoning the
interleaving that is the hallmark of SER.

\section{Concluding remarks}\label{sec:concl}

Algorithms SER and SERA are methods for link scheduling in WMNs. As such, and
unlike other methods for link scheduling, they are built around a set of
origin-to-destination paths and aim to provide as much throughput on these paths
as possible. From a mathematical perspective they are both related to providing
the nodes of a graph with an efficient multicoloring, in the sense discussed in
Section~\ref{sec:multic}. For SER this is strictly true, but for SERA the
defining characteristic of a multicoloring, that each node receives the same
number of colors, ceases to hold. As we demonstrated through our computational
results in Section~\ref{sec:results}, it is precisely this deviation from the
strict definition that allows SERA to surpass SER in terms of performance.

The functioning of both SER and SERA is supported by the use of the integer
parameter $B\ge 1$, which indicates how many buffering positions each WMN node
has to store in-transit packets for each of the paths that go through it.
Choosing $B=1$ suffices for SER because of its inherent property of alternating
interfering transmissions, but $B>1$ may in principle be needed for the
advantages of SERA to become manifest. In the simulations we conducted, however,
only rarely has this been the case, since on average increasing $B$ beyond $1$
provided no distinguishable improvement. In this regard, we find it important
for the reader to refer to Figure~\ref{fig5} once again. As we remarked upon
discussing that figure, profiting from a $B>1$ situation under SERA is largely
a matter of how uniformly interference gets distributed on the particular set of
paths at hand. Our results in Section~\ref{sec:results}, therefore, can safely
be assumed to have stemmed from circumstances that, on average, led to highly
uniformly distributed interference patterns.

The centerpiece of both SER and SERA is the undirected graph $G$, which embodies
a representation of all the interference affecting the various wireless links
represented by the graph's nodes. As we explained in Section~\ref{sec:transf},
the steps to building $G$ depend on how one assumes the communication and
interference radii to relate to each other, and also on which interference model
is adopted. We have given results for a specific set of assumptions, but clearly
there is nothing in either method precluding its use under any other
assumptions: all that needs to be done is construct $G$ accordingly.

Analyzing either method mathematically is a difficult enterprise, but since
their performance depends on the heuristic choice of an initial acyclic
orientation of $G$, any effort profitably spent in that direction will be
welcome. In addition to potentially better decisions regarding initial
conditions, further mathematical knowledge on SER or SERA may also come to
provide a deeper understanding of how upper bounds on $T(\mathcal{S})$ relate to
what is observed. As we mentioned in Section~\ref{sec:disc}, one such bound is
already known in the case of SER. Obtaining better bounds in this case, as well
as some bound in the case of SERA, remains open to further research.

Another issue that is open to further investigation is how to handle the
potential difficulties that SER may encounter in the face of a growing number of
nodes in $G$ \cite{mr92,mmz93}. These difficulties refer to the fact that, in
the worst case, the time required to detect the occurrence of the period may
grow exponentially with the square root of the number of nodes. They are
inherited by SERA, since it generalizes SER, and may require the development of
further heuristics if they pose a real problem in practice. In a related vein,
sometimes it may be the case that only the value of $T(\mathcal{S})$ is needed,
not $\mathcal{S}$ itself. Knowing the achievable throughput without requiring
knowledge of the schedule itself can be useful for evaluating WMN topologies or
routing algorithms for them.

Should this be the case, then it is possible to estimate $T(\mathcal{S})$ more
efficiently than using the full-fledged algorithms we gave. We can do this in
the case of SERA by recognizing that $T(\mathcal{S})$ is the limit, as
$t\to\infty$, of
\begin{equation}
T_t(\mathcal{S})=
\frac{\sum_{i\in T}m_i(\omega_0,t)}{t+1},
\end{equation}
where $m_i(\omega_0,t)$ is the total number of times node $i$ appears as a sink
in orientations $\omega_0,\omega_1,\ldots,\omega_t$. To see this, let $o(t)$
denote any function of $t$ such that $\lim_{t\to\infty}o(t)/t=0$. We then have
$\sum_{i\in T}m_i(\omega_0,t)=r(t)\sum_{i\in T}m_i(\omega_0)+o(t)$ and
$t+1=r(t)p(\omega_0)+o(t)$, where $r(t)$ is the number of times the period has
been repeated up to iteration $t$. The limit follows easily, and automatically
holds also for SER by straightforward extension. The streamlined version of
either algorithm consists simply of letting $t$ evolve either through a
sufficiently large value determined beforehand or until $T_t(\mathcal{S})$
becomes stable. Any of the two alternatives does away with the need to detect
the occurrence of the period.

We note, finally, that we have found the results given in Figure~\ref{fig10}
to be practically indistinguishable from those obtained through the strategy
outlined above for the computation of $T(\mathcal{S})$. We have verified this by
letting $T(\mathcal{S})=T_{t^+}(\mathcal{S})$, where $t^+$ is the least value of
$t$ for which
$\vert T_t(\mathcal{S})-T_{t-w}(\mathcal{S})\vert/T_{t-w}(\mathcal{S})\le 0.001$.
Here $w$ is a window parameter and in our experiments we used $w=\vert N\vert$.
As for this particular choice, it comes from realizing that in both SER and SERA
it takes at most $\vert N\vert-1$ iterations for a node of $G$ that is currently
not a sink to become one. This, in turn, comes from the fact that in each
iteration either algorithm necessarily decreases by $1$ the number of edges on a
longest directed path from any non-sink node to a sink. We can see that this is
true of SER by viewing its dynamics in terms of how the orientations' sink
decompositions evolve. We can see that it continues to hold in the case of SERA
because SERA never places a former sink $i$ into one of the sink-decomposition
sets that already contains a neighbor of $i$ in $G$ (cf.\ Figure~\ref{fig6}).

\subsection*{Acknowledgments}

We acknowledge partial support from CNPq, CAPES, a FAPERJ BBP grant, and a
scholarship grant from Universit\'{e} Pierre et Marie Curie.
All computational experiments were carried out on the Grid'5000 experimental
testbed, which is being developed under the INRIA ALADDIN development action
with support from CNRS, RENATER, and several universities as well as other
funding bodies (see \texttt{https://www.grid5000.fr}).

\bibliography{sera}
\bibliographystyle{plain}

\end{document}